\documentclass{article}

\usepackage{arxiv}

\usepackage[utf8]{inputenc}
\usepackage[T1]{fontenc}
\usepackage{graphicx}
\usepackage{booktabs}
\usepackage{tabularx}
\usepackage{multirow}
\usepackage{array}
\usepackage{longtable}
\usepackage{float}
\usepackage{caption}
\usepackage[dvipsnames]{xcolor}
\usepackage{tikz}
\usetikzlibrary{shapes.geometric, arrows.meta, positioning, fit, calc, backgrounds, decorations.pathreplacing}
\usepackage{enumitem}
\usepackage[hidelinks]{hyperref}
\usepackage{natbib}
\usepackage{amsmath}
\usepackage{amssymb}
\usepackage{url}

\definecolor{haifblue}{HTML}{2C5F8A}
\definecolor{tier1}{HTML}{4A90D9}
\definecolor{tier2}{HTML}{5BA85B}
\definecolor{tier3}{HTML}{E8A838}
\definecolor{tier4}{HTML}{D65D5D}
\definecolor{darkgray}{HTML}{444444}

\newcolumntype{L}[1]{>{\raggedright\arraybackslash}p{#1}}
\newcolumntype{C}[1]{>{\centering\arraybackslash}p{#1}}

\title{HAIF: A Human--AI Integration Framework\\for Hybrid Team Operations\\[0.3cm]
{\large\itshape An Operational Framework for Managing Collaborative Work Between Human Professionals and AI}}

\author{
  Marc Bara, Ph.D.\\
  ProjectWorkLab SL\\
  Barcelona, Spain\\
  \texttt{marc.bara@projectworklab.com}
}

\date{7th February 2026}

\begin{document}

\maketitle

\begin{abstract}
\noindent The rapid deployment of generative AI, copilots, and agentic systems in knowledge work has created an operational gap: while recent practitioner guides identify the challenges of human--AI collaboration, no existing work provides the operational protocols that connect delegation decisions to validation workflows to effort estimation within daily team practice. Agile, DevOps, MLOps, and AI governance frameworks each cover adjacent concerns but none models the hybrid team as a coherent delivery unit. This paper proposes the Human--AI Integration Framework (HAIF): a protocol-based, scalable operational system built around four core principles, a formal delegation decision model, tiered autonomy with quantifiable transition criteria, and feedback mechanisms designed to integrate into existing Scrum and Kanban workflows without requiring additional roles for small teams. The framework is developed following a Design Science Research methodology. HAIF explicitly addresses the central adoption paradox: the more capable AI becomes, the harder it is to justify the oversight the framework demands---and yet the greater the consequences of not providing it. The paper includes domain-specific validation checklists, adaptation guidance for non-software environments, and an examination of the framework's structural limitations---including the increasingly common pattern of continuous human--AI co-production that challenges the discrete delegation model. The framework is tool-agnostic and designed for iterative adoption. Empirical validation is identified as future work.

\vspace{0.5em}
\end{abstract}

\section{Introduction}
\label{sec:introduction}

Since 2023, generative AI has moved from experimental pilots to embedded operational tooling in organizations worldwide. Large language models are no longer confined to chatbot interfaces; they function as coding assistants, document drafters, data analysts, and decision-support systems \citep{Brynjolfsson2023, DellAcqua2023}. Agentic architectures---systems that execute multi-step tasks with limited supervision---have begun to appear in enterprise environments, further blurring the boundary between tool use and task delegation \citep{Wang2024, Xi2023}.

This creates a problem that is operational, not merely technical or ethical. When a non-human agent drafts a report, generates code, or synthesizes data, questions arise that no existing framework answers systematically: who owns the output? How is it validated? How do you estimate work when generation takes seconds but validation takes hours? And how do you prevent the gradual erosion of the very human expertise on which oversight depends?

Organizations are answering these questions daily through improvisation. The result is inconsistency in accountability, unpredictable quality, distorted planning, and a creeping dependency on systems whose failure modes are poorly understood. Several relevant bodies of knowledge exist---Agile for team coordination, DevOps for system reliability, MLOps for model lifecycle, AI governance for risk and compliance, human-in-the-loop for decision oversight---but none addresses the management of hybrid human--AI teams as productive units.

This paper proposes the Human--AI Integration Framework (HAIF) to fill this gap. HAIF is not a meta-framework or a set of principles to hang on a wall. It is a protocol-based operational system designed to integrate into existing Agile and Kanban practices, scale from individual practitioners to enterprise PMOs, and confront rather than avoid the hard questions that hybrid work introduces.

\textbf{Research objective.} To define a formal, scalable operational framework for hybrid human--AI teams, grounded in the identification of specific operational challenges that current frameworks leave unresolved, with explicit decision models rather than abstract principles. The framework is developed following a Design Science Research methodology (Section~\ref{sec:methodology}). Empirical validation through field studies is identified as future work.

\section{Research Methodology}
\label{sec:methodology}

This paper follows the Design Science Research (DSR) methodology as formalized by \citet{Hevner2004} and operationalized through the process model of \citet{Peffers2007}. DSR is appropriate when the research contribution is an artifact---a framework, model, method, or instantiation---designed to address an identified organizational problem. Unlike behavioral research, which seeks to describe and explain phenomena, design science seeks to create and evaluate artifacts that extend human and organizational capabilities \citep{Hevner2004}.

\subsection{Methodological Fit}

The problem addressed by this paper---the absence of an operational framework for hybrid human--AI teams---is a design problem: it requires the creation of a new artifact (an operational framework) rather than the explanation of an observed phenomenon. DSR is the established methodology for such contributions in Information Systems research \citep{Gregor2013} and has been successfully applied to the development of project management frameworks, governance models, and organizational design artifacts.

\subsection{DSR Process}

Following \citet{Peffers2007}, the research proceeds through six activities:

\begin{enumerate}[leftmargin=2cm]
\item \textbf{Problem identification and motivation} (Sections~\ref{sec:related}--\ref{sec:problem}). The operational gap is identified through a critical review of existing frameworks (Agile, DevOps, MLOps, AI governance, HITL) and the characterization of five specific challenges that hybrid teams face.

\item \textbf{Definition of objectives for a solution} (Section~\ref{sec:problem}, \S3.6). The framework must address delegation governance, accountability, quality assurance, estimation under asymmetric effort, and cognitive dependency---while being lightweight enough for adoption under delivery pressure.

\item \textbf{Design and development} (Sections~\ref{sec:principles}--\ref{sec:structure}). The HAIF framework is designed as a protocol-based system with four principles, a formal delegation decision model, tiered autonomy with quantifiable transition criteria, and integration mechanisms for existing workflows.

\item \textbf{Demonstration} (Section~\ref{sec:workflow}). The framework is demonstrated through an illustrative scenario based on a product development team operating under realistic conditions, including process violations, quality surprises, and delivery pressure.

\item \textbf{Evaluation}. This activity is partially addressed through analytical evaluation: comparison with existing frameworks (Section~\ref{sec:comparison}), examination of hard questions and structural limitations (Section~\ref{sec:hard}), and expert review. Full empirical evaluation through field studies and controlled experiments is identified as future work (Section~\ref{sec:limitations}).

\item \textbf{Communication}. This paper constitutes the communication activity.
\end{enumerate}

\subsection{Knowledge Contribution}

Following the DSR knowledge contribution framework of \citet{Gregor2013}, HAIF represents a Level~2 contribution: a \textit{nascent design theory}---an artifact that addresses a new or previously unsolved problem (hybrid team operations) with a novel solution (protocol-based operational integration of AI agents into structured team workflows). The contribution is prescriptive: it defines how hybrid teams \textit{should} be organized to address identified operational challenges, based on theoretical grounding in human factors, automation research, and project management practice.

\subsection{Design Grounding}

The framework's design draws on three bodies of established knowledge:

\begin{itemize}[leftmargin=1.5cm]
\item \textit{Human factors and automation research.} The autonomy tier model extends the levels-of-automation taxonomy of \citet{Parasuraman2000}, adapted from supervisory control to team workflow management. The cognitive dependency construct draws on automation complacency research \citep{Parasuraman2010, Bainbridge1983, Goddard2012}.

\item \textit{Agile and lean management.} The protocol-layer integration model builds on the architectural principles of Scrum \citep{Schwaber2020} and Kanban \citep{Anderson2010}, extending their ceremonies and artifacts rather than replacing them.

\item \textit{Quality management.} The tiered validation approach adapts acceptance sampling and statistical quality control principles \citep{Montgomery2019} to the specific characteristics of AI-generated outputs.
\end{itemize}

\subsection{Limitations of the Methodology}

DSR explicitly permits the publication of artifacts prior to full empirical evaluation, provided that the evaluation gap is acknowledged and future evaluation plans are specified \citep{Hevner2004, Peffers2007}. This paper completes DSR activities 1--4 and partially addresses activity~5 through analytical evaluation. The absence of field evaluation is a limitation, not a methodological deficiency---it defines the boundary of the current contribution and the agenda for subsequent research.

\section{Related Work}
\label{sec:related}

This section reviews existing frameworks with a specific lens: what does each assume about who performs the work, and what breaks when that assumption no longer holds?

\subsection{Agile and Scrum}

Agile methodologies \citep{Beck2001}, operationalized through Scrum \citep{Schwaber2020} and Kanban \citep{Anderson2010}, assume all team members are human agents capable of self-organization, mutual communication, shared understanding, and reflective learning. Velocity is calibrated to human effort. Retrospectives assume metacognition. Story points presume roughly comparable cognitive profiles among estimators.

When AI performs a portion of sprint work, these assumptions break in specific ways. Velocity becomes unreliable because the same ``story'' may take 2~hours with AI or 16 without---but the validation overhead differs. Retrospectives cannot include the AI agent as a participant capable of behavioral change. The Definition of Done implicitly assumes human-quality peer review, which is qualitatively different from reviewing machine-generated output. Scrum does not model non-human contributors, and its ceremonies provide no mechanism for managing delegation boundaries, validation protocols, or the specific failure modes of AI-generated work.

\textbf{Covers:} iterative cadence, team coordination, transparent inspection, adaptive planning. \textbf{Does not cover:} non-human contributors, asymmetric effort profiles, machine-output validation, cognitive dependency.

\subsection{DevOps and Site Reliability Engineering}

DevOps \citep{Kim2016} and SRE \citep{Beyer2016} manage the intersection of development and operations through automation, CI/CD, monitoring, error budgets, and SLOs. Both excel at ensuring systems run correctly.

The operational question in DevOps is \textit{``is the system working?''}---not \textit{``is the system's productive output meeting the same quality and accountability standards as human-produced work?''} DevOps monitors uptime and latency; it does not monitor whether an AI-generated report is factually accurate or contextually appropriate. The conceptual infrastructure of SRE---particularly error budgets and SLO-driven management---is relevant to HAIF and is explicitly adapted, but SRE itself does not address delegation of cognitive tasks or governance of generated outputs.

\textbf{Covers:} system reliability, deployment, incident response, monitoring. \textbf{Does not cover:} output quality in work contexts, delegation governance, human skill preservation.

\subsection{MLOps}

MLOps \citep{Sculley2015, Amershi2019, Kreuzberger2023} manages the lifecycle of machine learning models: data versioning, training, deployment, drift detection, retraining. Its focus is the technical artifact---the model---and its behavior in production.

The gap is between \textit{model lifecycle} and \textit{work lifecycle}. MLOps ensures a model performs within technical parameters. It does not prescribe who reviews the model's outputs in a work context, what quality threshold applies, how the output integrates into a team's delivery, or who is accountable when a model-generated artifact becomes part of a client deliverable. HAIF operates downstream of MLOps, governing what happens after the model produces output.

\textbf{Covers:} model training, deployment, monitoring, drift, retraining. \textbf{Does not cover:} output governance in team workflows, validation in delivery context, accountability assignment.

\subsection{AI Governance and Responsible AI}

AI governance frameworks---EU AI Act \citep{EUAI2024}, NIST AI RMF \citep{NIST2023}, organizational Responsible AI programs \citep{Microsoft2022, Google2023}---address ethical, legal, and risk dimensions. They prescribe fairness assessments, transparency requirements, human oversight obligations, and risk classifications.

Their limitation is one of abstraction level. A governance framework may require human oversight of high-risk AI outputs, but it does not specify how that oversight integrates into a sprint, who performs it, how it is estimated, what the review criteria are, or how exceptions are handled. Governance defines constraints; it does not define workflows.

\textbf{Covers:} compliance, risk classification, fairness, organizational policy. \textbf{Does not cover:} daily operational workflow, validation protocols, team-level implementation.

\subsection{Human-in-the-Loop}

HITL \citep{Amershi2014, Monarch2021} ensures human decision authority at critical points. HITL provides useful conceptual vocabulary, particularly the levels-of-automation taxonomy from \citet{Parasuraman2000}.

Two limitations restrict its applicability to hybrid teams. First, HITL addresses individual decision points, not sustained collaborative workflows. Second, HITL assumes a relatively static allocation of human and machine roles, whereas operational hybrid teams require dynamic reallocation based on evolving task characteristics, risk, and capability. HAIF draws on the Parasuraman et~al. taxonomy but extends it with operational transition criteria and integration into team workflows rather than individual decision gates.

\textbf{Covers:} decision-point oversight, autonomy level conceptualization. \textbf{Does not cover:} sustained workflow integration, dynamic reallocation, team-level protocols.

\subsection{AI Integration in Agile Practice}

A growing body of practitioner-oriented work addresses the intersection of AI and agile workflows, particularly Scrum. This literature identifies many of the same operational challenges that motivate HAIF but stops short of prescriptive, protocol-level solutions.

The \textit{Scrum Guide Expansion Pack: AI and Scrum} \citep{Jocham2026} provides the most comprehensive treatment to date. Co-authored with Scrum co-creator Jeff Sutherland, it synthesizes evidence on AI's impact on development teams, documents the productivity paradox (senior developers experiencing $\sim$19\% performance decreases on complex AI-assisted tasks due to verification overhead), identifies the ``two-speed IT'' problem (AI-accelerated generation outpacing organizational validation capacity), and recommends ``manual days'' for skill preservation. It articulates \textit{what} teams should attend to---transparency of AI-generated work, human accountability, quality discipline, validation capacity---but leaves \textit{how} to implement these at the protocol level as an exercise for each team.

The \textit{AI Teammate Framework} \citep{Fernandes2025}, published by Scrum.org, treats AI integration as a four-step team management process: Model Management (selecting AI tools), Context Management (onboarding AI with domain knowledge), Prompt Engineering (optimizing interactions), and Governance Management (evaluating performance). Its focus is on how teams \textit{use} AI effectively---prompt quality, context provision, cost management---rather than on how teams \textit{govern the outputs} AI produces within delivery workflows. It does not propose autonomy tiers, validation protocols, or effort estimation models for hybrid work.

\citet{Isobe2025} proposes a three-tiered estimation scheme for AI-assisted Scrum: ``Zero-Point'' stories (fully automated), ``Standard'' stories (human-led), and ``Review \& Integration'' stories (hybrid). This directly addresses effort asymmetry but without a formal delegation model, validation criteria, or tier transition mechanisms. The XP2025 workshop research roadmap \citep{XP2025Roadmap} confirms these gaps as community-recognized open problems, identifying tool fragmentation, governance uncertainty, and skills gaps as the three dominant themes requiring structured research response.

Collectively, this literature establishes consensus on the problem space: AI disrupts Scrum's assumptions about effort, quality, and accountability. It also converges on \textit{what} teams should attend to---transparency, human ownership, validation discipline, skill preservation. What no existing work provides is the operational machinery: a formal model for deciding \textit{what} to delegate at \textit{what} autonomy level, \textit{how much} validation to budget, \textit{when} to demote a delegation, and \textit{how} to detect competence erosion before it manifests as failure. The gap is not in problem identification but in protocol specification. HAIF addresses this gap directly, translating the strategic recommendations of \citet{Jocham2026} and \citet{Fernandes2025} into concrete, sprint-level mechanisms with explicit decision criteria, quantified validation overhead, and reversible autonomy tiers.

\subsection{Summary: The Consolidated Gap}

Table~\ref{tab:gaps} summarizes coverage across frameworks.

\begin{table}[H]
\centering
\caption{Coverage analysis of existing frameworks against hybrid team operational needs.}
\label{tab:gaps}
\small
\renewcommand{\arraystretch}{1.3}
\begin{tabularx}{\textwidth}{L{3cm} C{1.1cm} C{1.1cm} C{1.1cm} C{1.1cm} C{1.1cm} C{1.1cm}}
\toprule
\textbf{Operational Need} & \textbf{Scrum} & \textbf{DevOps} & \textbf{MLOps} & \textbf{AI Gov.} & \textbf{HITL} & \textbf{HAIF} \\
\midrule
Task delegation to AI          & --  & --  & --  & $\sim$ & $\sim$ & \checkmark \\
Accountability for outputs     & --  & --  & --  & \checkmark & $\sim$ & \checkmark \\
QA of machine-generated work   & --  & $\sim$ & $\sim$ & $\sim$ & $\sim$ & \checkmark \\
Asymmetric effort estimation   & --  & --  & --  & --  & --  & \checkmark \\
Cognitive dependency prevention & -- & --  & --  & --  & --  & \checkmark \\
Tiered autonomy + transitions  & --  & --  & --  & $\sim$ & \checkmark & \checkmark \\
Integrated workflow protocols  & \checkmark & \checkmark & \checkmark & -- & -- & \checkmark \\
Provenance and traceability    & --  & $\sim$ & \checkmark & \checkmark & -- & \checkmark \\
\bottomrule
\end{tabularx}
\vspace{0.3em}
{\footnotesize \checkmark\ = explicitly addressed; $\sim$ = partially; -- = not addressed.}
\end{table}

No existing framework provides a unified operational model for teams where AI agents perform substantive work alongside humans. Each covers a necessary adjacent concern; none covers the operational core of hybrid team management. Recent practitioner guides \citep{Jocham2026, Fernandes2025} have correctly identified the key tensions---effort asymmetry, accountability, skill erosion, validation capacity---but articulate them as strategic recommendations rather than operational protocols. Recent theoretical work on hybrid intelligence teams \citep{Eccles2025} has begun to articulate the cognitive and organizational foundations---shared mental models, transactive memory, epistemic safety---that underpin effective human--AI collaboration. HAIF bridges these layers by providing the operational protocols, delegation governance, and workflow integration mechanisms that translate strategic guidance and theoretical principles into daily practice.

\section{Problem Definition}
\label{sec:problem}

This section characterizes five operational challenges that hybrid teams face, examined through the lens of what specifically breaks in practice and why ad~hoc solutions are insufficient.

\subsection{The Delegation Problem}

In human teams, delegation relies on shared context, the delegate's capacity to recognize ambiguity, and implicit professional norms. Delegating to an AI agent is fundamentally different: the agent cannot recognize when a task is underspecified in ways that matter, does not hold professional judgment, and may produce outputs that are fluent but wrong \citep{Noy2023}. The boundary of what can be safely delegated is not static---it depends on the specific AI system's current capability, the task's risk profile, the availability of qualified reviewers, and the organizational context.

\textbf{What breaks:} Teams either under-delegate (wasting AI potential) or over-delegate (producing unreviewed outputs that enter delivery pipelines). Both failure modes are invisible until consequences surface.

\subsection{The Accountability Gap}

When an AI generates an output incorporated into a deliverable, accountability becomes ambiguous. The AI system cannot be held accountable in any meaningful organizational or legal sense. Current practice resolves this through informal norms---meaning accountability is ambiguous precisely when it matters most: when something goes wrong.

\textbf{What breaks:} Post-mortem analyses cannot identify clear ownership. Teams develop blame-avoidance behaviors rather than learning behaviors.

\subsection{The Validation Paradox}

Reviewing AI-generated work requires detecting confident-sounding errors, hallucinated references, statistical fabrications, and logical inconsistencies masked by fluent prose \citep{Ji2023, Huang2023}. The paradox: organizations adopt AI to reduce the workload of their most skilled people---but validating AI outputs effectively requires \textit{exactly those skills}. The cheaper the generation, the more expensive the validation per unit of genuine quality assurance.

\textbf{What breaks:} Organizations apply the same review standards to AI and human work. Since AI output looks polished, it receives \textit{less} scrutiny---the opposite of what is needed.

\subsection{The Effort Asymmetry}

AI can generate a 20-page report in seconds. Validating it takes hours. Traditional estimation methods are calibrated to human production speed. When generation is near-instant, the effort profile inverts: the bulk of work shifts to specification and validation rather than execution.

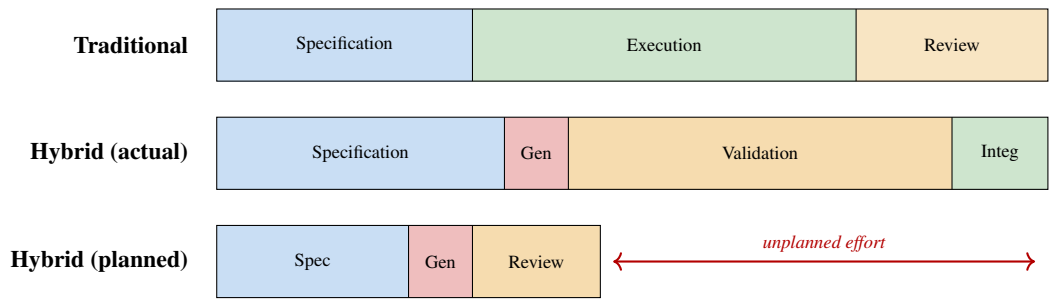
\begin{figure}[H]
\centering
\begin{tikzpicture}[x=0.85cm, y=1.2cm]
  \node[anchor=east, font=\small\bfseries] at (-0.3, 2) {Traditional};
  \fill[tier1!30] (0,1.6) rectangle (4,2.4);
  \fill[tier2!30] (4,1.6) rectangle (10,2.4);
  \fill[tier3!30] (10,1.6) rectangle (13,2.4);
  \draw (0,1.6) rectangle (13,2.4);
  \draw (4,1.6) -- (4,2.4);
  \draw (10,1.6) -- (10,2.4);
  \node[font=\scriptsize] at (2,2) {Specification};
  \node[font=\scriptsize] at (7,2) {Execution};
  \node[font=\scriptsize] at (11.5,2) {Review};

  \node[anchor=east, font=\small\bfseries] at (-0.3, 0.8) {Hybrid (actual)};
  \fill[tier1!30] (0,0.4) rectangle (4.5,1.2);
  \fill[tier4!40] (4.5,0.4) rectangle (5.5,1.2);
  \fill[tier3!40] (5.5,0.4) rectangle (11.5,1.2);
  \fill[tier2!30] (11.5,0.4) rectangle (13,1.2);
  \draw (0,0.4) rectangle (13,1.2);
  \draw (4.5,0.4) -- (4.5,1.2);
  \draw (5.5,0.4) -- (5.5,1.2);
  \draw (11.5,0.4) -- (11.5,1.2);
  \node[font=\scriptsize] at (2.25,0.8) {Specification};
  \node[font=\scriptsize] at (5,0.8) {Gen};
  \node[font=\scriptsize] at (8.5,0.8) {Validation};
  \node[font=\scriptsize] at (12.25,0.8) {Integ};

  \node[anchor=east, font=\small\bfseries] at (-0.3, -0.4) {Hybrid (planned)};
  \fill[tier1!30] (0,-0.8) rectangle (3,0);
  \fill[tier4!40] (3,-0.8) rectangle (4,0);
  \fill[tier3!40] (4,-0.8) rectangle (6,0);
  \draw (0,-0.8) rectangle (6,0);
  \draw (3,-0.8) -- (3,0);
  \draw (4,-0.8) -- (4,0);
  \node[font=\scriptsize] at (1.5,-0.4) {Spec};
  \node[font=\scriptsize] at (3.5,-0.4) {Gen};
  \node[font=\scriptsize] at (5,-0.4) {Review};

  \draw[<->, thick, red!70!black] (6.2,-0.4) -- (12.8,-0.4);
  \node[font=\scriptsize\itshape, red!70!black, above] at (9.5,-0.4) {unplanned effort};

\end{tikzpicture}
\caption{Task effort distribution: traditional vs.\ hybrid work. Teams systematically underestimate validation effort, planning for the ``perceived'' distribution while actual effort follows the middle pattern.}
\label{fig:effort-distribution}
\end{figure}

\textbf{What breaks:} Teams overcommit based on perceived AI productivity, then discover that validation backlogs consume the time saved.

\subsection{The Dependency Trap}

As professionals delegate cognitive tasks to AI, their ability to perform those tasks independently degrades. This is documented in aviation \citep{Parasuraman2010}, healthcare \citep{Goddard2012}, and is emerging in knowledge work \citep{DellAcqua2023}. \citet{Bainbridge1983} identified this four decades ago: the more reliable the system, the less practiced the operator, the less capable they are of intervening when the system fails.

\textbf{What breaks:} Senior team members initially catch AI errors effectively. Over months of reduced practice, their detection rate declines silently.

\subsection{The Adoption Paradox}
\label{subsec:paradox}

These five challenges share a meta-challenge:

\textit{The more capable AI becomes, the harder it is to justify the operational discipline the framework demands---and yet the more necessary that discipline becomes.}

Organizations under delivery pressure resist adding validation overhead precisely when AI appears to be working well. The framework must therefore produce visible value---better estimates, fewer post-delivery defects, clear accountability---sufficient to justify its adoption cost. This tension between discipline and pressure is the central design constraint of HAIF.

\section{Core Principles}
\label{sec:principles}

\begin{figure}[H]
\centering
\begin{tikzpicture}[
    >=Stealth,
    principle/.style={rectangle, rounded corners=4pt, draw=haifblue, fill=haifblue!8, thick, minimum height=0.95cm, text width=7cm, align=left, font=\small},
    numcircle/.style={circle, fill=haifblue, text=white, font=\small\bfseries, minimum size=0.7cm, inner sep=0},
]
\foreach \i/\txt/\sub in {
    1/{Named Human Ownership}/{Every AI output has an accountable person},
    2/{Governed, Reversible Delegation}/{Explicit, tiered, visible, demotable},
    3/{Proportional, Planned Validation}/{Budgeted, measured, tier-specific},
    4/{Active Competence Maintenance}/{Periodic human-only cycles, skill calibration}
} {
    \pgfmathsetmacro{\ypos}{5.0 - \i * 1.35}
    \node[numcircle] (n\i) at (0, \ypos) {\i};
    \node[principle, anchor=west] (p\i) at (0.65, \ypos) {\textbf{\txt}\\\scriptsize\textit{\sub}};
}
\draw[decorate, decoration={brace, amplitude=8pt}, thick, haifblue!60]
  ([xshift=6pt]p1.north east) -- ([xshift=6pt]p2.south east)
  node[midway, right=10pt, font=\small\itshape, text=haifblue] {Governance};
\draw[decorate, decoration={brace, amplitude=8pt}, thick, haifblue!60]
  ([xshift=6pt]p3.north east) -- ([xshift=6pt]p4.south east)
  node[midway, right=10pt, font=\small\itshape, text=haifblue] {Sustainability};
\end{tikzpicture}
\caption{The four core principles of HAIF.}
\label{fig:principles}
\end{figure}
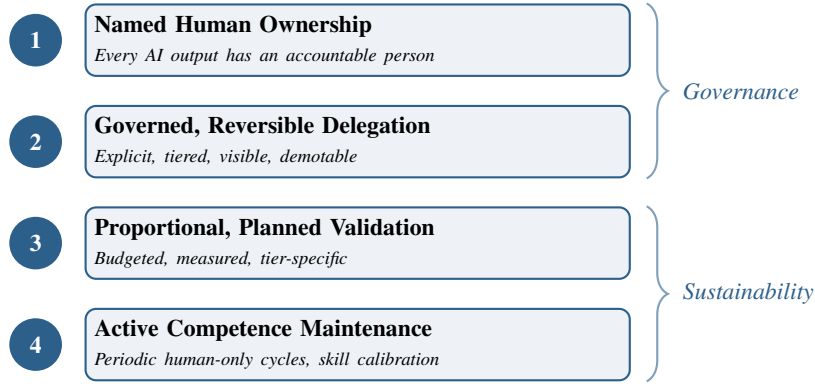

HAIF is grounded in four principles, each with direct operational consequences. Figure~\ref{fig:principles} provides an overview.

\subsection{Principle 1: Every AI Output Has a Named Human Owner}

No AI-generated output may enter a delivery pipeline, cross a team boundary, or inform an organizational decision without a named human being accountable for its content. This is absolute and does not vary with autonomy tier.

\textbf{Operational consequence:} Before any AI-delegated task begins, the delegation protocol requires explicit assignment of a human owner. This person does not need to review every output personally---at higher autonomy tiers, they may verify that monitoring systems functioned correctly---but their name is attached and they accept accountability.

\textbf{Why this cannot be softened:} If accountability resolves to a ``process'' rather than a person, no one has sufficient incentive to maintain validation quality.

\subsection{Principle 2: Delegation Is a Governed, Reversible Act}

Delegating work to an AI agent is not tool use---it is an operational decision with implications for quality, accountability, and risk. Every delegation must be explicit (visible in planning artifacts), governed (assigned an autonomy tier through the decision model), and reversible (reassignable to human execution without friction or stigma).

\textbf{Operational consequence:} The team maintains a \textit{delegation registry}---a living document recording what is delegated, at what tier, with what rationale, and with what historical performance data. Implicit delegation is a framework violation.

\textbf{Why reversibility matters:} Organizations under productivity pressure ratchet delegation upward. Reversibility without stigma ensures tier demotions are routine operational decisions.

\subsection{Principle 3: Validation Is Proportional, Planned, and Measured}

All AI-generated outputs are subject to validation, but the nature of validation varies by autonomy tier. Validation is a first-class operational activity that consumes planned capacity, follows defined protocols, and produces measurable data.

\textbf{Operational consequence:} Teams budget validation capacity explicitly. Validation findings are recorded: error rates, error types, review time, false acceptance rates. This data drives tier transitions and provides the evidence base for framework adaptation.

\subsection{Principle 4: Human Competence Is Actively Maintained}

The framework's oversight mechanisms depend on human expertise that must be deliberately preserved through periodic human-only execution, structured evaluation, and professional development not contingent on AI availability.

\textbf{Operational consequence:} For any task type delegated at Tier~2 or above, the team schedules periodic human-only cycles---not as punishment, but as calibration. On the board, these items carry the \textit{AI-restricted} classification.

\textbf{Why this is operational, not HR:} Skill preservation is exercised weekly (in validation), not annually (in a training session).

\section{The Delegation Decision Model}
\label{sec:structure}

This section provides a concrete decision model for delegation, tier assignment, and tier transitions.

\textbf{Scope of classification.} In contemporary work environments, virtually all knowledge workers use AI to some degree---from autocomplete and search augmentation to full output generation. HAIF therefore requires that \textit{every} backlog item receives a tier classification. Tier~1 (Assisted) is the default: the human drives the work, AI supports. Most items in most sprints will be Tier~1. Higher tiers (2--4) apply when AI produces output that requires specific validation protocols. The classification \textbf{AI-restricted} applies when the team explicitly prohibits AI involvement---either because consequence severity demands unmediated human judgment (e.g., security audits, legal review) or because competence maintenance (Principle~4) requires periodic human-only execution. Items without any AI involvement do not exist as a separate category; in practice, they are Tier~1 items where AI assistance was minimal or absent. The key design choice is that no item passes through planning unclassified---this makes AI use visible and prevents unexamined delegation.

\subsection{Decision Inputs}

For any candidate task, four assessments are required:

\textbf{Structuredness (S):} How well-defined are inputs, constraints, and expected output? Low / Medium / High.

\textbf{Verifiability (V):} How feasible is output verification? Low (deep expertise, subjective) / Medium (verifiable with effort) / High (objectively verifiable, automatable).

\textbf{Consequence of Error (C):} What if the output contains errors? Low (internal, correctable) / Medium (external, moderate cost) / High (legal, financial, safety, reputational).

\textbf{AI Demonstrated Capability (D):} Observed quality on this specific task type. Unproven / Emerging / Established / Mature.

\subsection{Decision Matrix}

Table~\ref{tab:matrix} maps input combinations to autonomy tiers.

\begin{table}[H]
\centering
\caption{Delegation decision matrix: input combinations to default autonomy tier.}
\label{tab:matrix}
\small
\renewcommand{\arraystretch}{1.3}
\begin{tabular}{L{4.5cm} C{2.2cm} C{2.2cm} C{2.2cm}}
\toprule
& \textbf{C: Low} & \textbf{C: Medium} & \textbf{C: High} \\
\midrule
S: High, V: High, D: Mature      & Tier 4 & Tier 3 & Tier 2 \\
S: High, V: High, D: Established & Tier 3 & Tier 2 & Tier 2 \\
S: High, V: Med, D: Established  & Tier 3 & Tier 2 & Tier 1 \\
S: Med, V: Med, D: Established   & Tier 2 & Tier 2 & Tier 1 \\
S: Med, V: Med, D: Emerging      & Tier 2 & Tier 1 & AI-restricted \\
S: Low \textit{or} V: Low        & Tier 1 & Tier 1 & AI-restricted \\
D: Unproven                      & Tier 1 (pilot) & Tier 1 (pilot) & AI-restricted \\
\bottomrule
\end{tabular}
\end{table}

\textbf{Key rule:} D can only advance through evidence accumulated at lower tiers. An AI system that has never performed a task type begins at Unproven regardless of its general capabilities.

\subsection{Tier Transition Criteria}

\textbf{Promotion} (upward): Minimum \textit{n} cycles at current tier (recommended: 3 for T1$\to$T2, 5 for T2$\to$T3, 8 for T3$\to$T4); error rate below tier threshold; no critical errors in the evaluation period; validation function confirms the proposed tier's protocol is feasible and resourced.

\textbf{Demotion} (downward): Immediate upon critical error detection; immediate upon error rate exceeding threshold for two consecutive cycles; immediate upon determination that validation capacity is insufficient; triggerable by any team member without formal approval.

\textbf{Asymmetry by design:} Promotion is slow and evidence-based. Demotion is fast and low-friction. This counteracts premature over-delegation.

\subsection{The Four Autonomy Tiers}

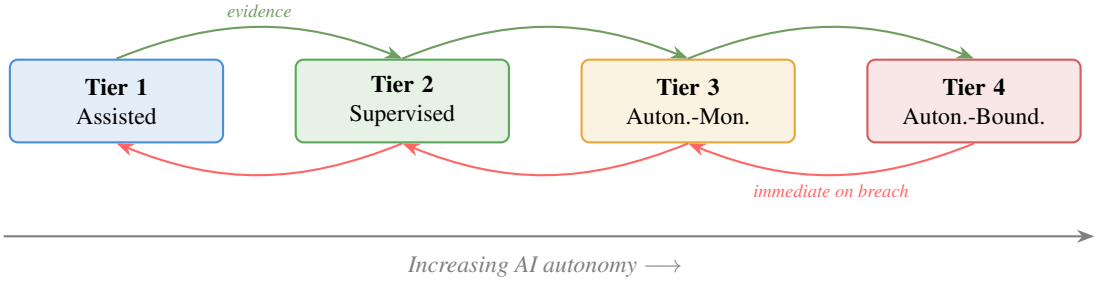
\begin{figure}[H]
\centering
\begin{tikzpicture}[
    >=Stealth,
    tierbox/.style={rectangle, rounded corners=3pt, thick, minimum height=1.1cm, text width=2.6cm, align=center, font=\small},
]
\node[tierbox, draw=tier1, fill=tier1!15] (t1) at (0,0) {\textbf{Tier 1}\\Assisted};
\node[tierbox, draw=tier2, fill=tier2!15] (t2) at (3.8,0) {\textbf{Tier 2}\\Supervised};
\node[tierbox, draw=tier3, fill=tier3!15] (t3) at (7.6,0) {\textbf{Tier 3}\\Auton.-Mon.};
\node[tierbox, draw=tier4, fill=tier4!15] (t4) at (11.4,0) {\textbf{Tier 4}\\Auton.-Bound.};
\draw[->, thick, OliveGreen!70] (t1.north) to[bend left=22] node[above, font=\scriptsize\itshape] {evidence} (t2.north);
\draw[->, thick, OliveGreen!70] (t2.north) to[bend left=22] (t3.north);
\draw[->, thick, OliveGreen!70] (t3.north) to[bend left=22] (t4.north);
\draw[->, thick, red!60] (t4.south) to[bend left=22] node[below, font=\scriptsize\itshape] {immediate on breach} (t3.south);
\draw[->, thick, red!60] (t3.south) to[bend left=22] (t2.south);
\draw[->, thick, red!60] (t2.south) to[bend left=22] (t1.south);
\draw[->, thick, gray] (-1.5, -1.8) -- (13, -1.8);
\node[font=\small\itshape, gray] at (5.7, -2.2) {Increasing AI autonomy $\longrightarrow$};
\end{tikzpicture}
\caption{Tier progression: slow promotion (green, evidence-based), fast demotion (red, immediate on quality breach).}
\label{fig:tiers}
\end{figure}

Table~\ref{tab:tiers} specifies each tier. Figure~\ref{fig:tiers} illustrates the progression model.

\begin{table}[H]
\centering
\caption{HAIF autonomy tier specifications.}
\label{tab:tiers}
\small
\renewcommand{\arraystretch}{1.3}
\begin{tabularx}{\textwidth}{C{0.6cm} L{2cm} L{3cm} L{3cm} L{3.5cm}}
\toprule
\textbf{Tier} & \textbf{Name} & \textbf{AI Role} & \textbf{Validation} & \textbf{Planning} \\
\midrule
\textbf{1} & Assisted & Supports human execution & Inherent in human process & Standard human estimate (AI acceleration varies by task) \\
\textbf{2} & Supervised & Produces output; human reviews before delivery & Full review, domain checklist (Appendix~A) & Spec (15--30\%) + Gen + Validation (30--60\%) \\
\textbf{3} & Auton.-Monitored & Produces output; post-hoc sampling & Sampling at \textit{p}\% + automated checks & Monitoring + sampling + exception handling \\
\textbf{4} & Auton.-Bounded & Independent within parameters; exception-based & Boundary monitoring + periodic audit & Boundary maintenance + audit + exceptions \\
\bottomrule
\end{tabularx}
\end{table}

For Tier~3 sampling rates: teams without historical data should begin at 20\%, adjusting downward as error baselines stabilize. Classical statistical quality control \citep{Montgomery2019} provides rigorous methods for determining sampling rates given acceptable quality levels and lot sizes; teams with sufficient data should adopt these. The key requirement is that the rate be explicit, documented, and reviewed each cycle.

\section{Framework Integration with Agile Workflows}
\label{sec:integration}

HAIF is a \textbf{protocol layer} that integrates into existing ceremonies and artifacts. It adds zero roles in small teams.

\subsection{Scaling Model}

Table~\ref{tab:scaling} shows how framework functions consolidate by team size.

\begin{table}[H]
\centering
\caption{HAIF scaling: function consolidation by team size.}
\label{tab:scaling}
\small
\renewcommand{\arraystretch}{1.3}
\begin{tabularx}{\textwidth}{L{2.5cm} L{3.5cm} L{3.5cm} L{3.5cm}}
\toprule
\textbf{Function} & \textbf{Solo/Pair (1--2)} & \textbf{Small Team (3--7)} & \textbf{Dept./PMO (8+)} \\
\midrule
HWO function & Practitioner & SM or Tech Lead & Dedicated role \\
Validation & Self-review + checklist & Peer review rotation & Validation Lead \\
Integration & Personal DoD & Team DoD & Integration Steward \\
Registry & Personal log & Board metadata/tags & Formal registry \\
Skill maint. & Self-scheduled blocks & Team rotation cycles & Programmatic plan \\
\bottomrule
\end{tabularx}
\end{table}

\subsection{Integration with Scrum Ceremonies}

\textbf{Sprint Planning:} Every backlog item carries a tier classification (Tier~1--4 or AI-restricted), a human owner, and an effort estimate that includes validation. Estimation includes full lifecycle. Validation capacity is explicitly budgeted---if the team cannot resource the implied validation, the plan is adjusted, not the validation.

\textbf{Daily Standup:} No structural change. Team members report on AI-delegated tasks they own. AI agents do not participate.

\textbf{Sprint Review:} Deliverables presented with provenance awareness. HWO function reports briefly on hybrid health: tier distribution, error rates.

\textbf{Sprint Retrospective:} Hybrid-specific questions: Were tier assignments accurate? Was validation effort correctly estimated? Quality surprises? Competence erosion signals? Tier adjustments are decided here and documented in the registry.

\subsection{Extended Definition of Done}

A work item involving AI-generated content is not done unless: (1)~output validated per tier protocol; (2)~provenance recorded; (3)~human owner confirmed; (4)~integration verified; (5)~deficiencies resolved or documented as accepted risk.

\subsection{Adaptation Beyond Scrum}
\label{subsec:beyond}

\textbf{Kanban.} Delegation classification occurs at commitment point (entering ``In Progress''). Tier recorded as card metadata. Validation triggered by WIP state transitions---items cannot move to ``Done'' without tier-specific protocol. Hybrid flow review biweekly, examining delegation patterns and error rates on rolling windows.

\textbf{Non-software knowledge work.} HAIF's protocols are domain-agnostic in structure but require domain-specific validation criteria. For consulting, ``validation'' means checking claims, analytical coherence, and client context. For marketing: brand voice, factual accuracy, strategic alignment. For legal: cross-referencing regulation and precedent. The framework provides the \textit{when, who, and how rigorously}; the team defines the \textit{what to check}. Appendix~A provides starter checklists.

\textbf{Research and creative environments.} Tier~1 (Assisted) is the natural default---the human drives the process, using AI to accelerate sub-tasks. Higher tiers apply only to well-characterized sub-tasks (formatting, citation management, data visualization). The framework's value here is making AI use visible and ensuring human creative judgment remains the driving force.

\section{Operational Workflow: An Illustrative Scenario}
\label{sec:workflow}

This section demonstrates HAIF through a realistic scenario: a six-person product team (PO, SM, four developers) building a B2B SaaS platform. Sprint length: two weeks. AI use: four months, no formal framework.

\subsection{Planning}

The SM, in HWO function, leads classification:

\begin{table}[H]
\centering
\small
\renewcommand{\arraystretch}{1.2}
\begin{tabularx}{\textwidth}{L{3.5cm} L{2cm} L{2cm} L{5cm}}
\toprule
\textbf{Item} & \textbf{Type} & \textbf{Tier} & \textbf{Rationale} \\
\midrule
API endpoints (3) & Code gen. & Tier 2 & D: Established; V: High; C: Medium \\
Unit tests (5) & Code gen. & Tier 3 & D: Mature (4 months); V: High; C: Low \\
User docs update & Doc. gen. & Tier 2 & D: Emerging; V: Medium; C: Medium \\
Security review & Analysis & AI-restricted & C: High; S: Low; deep judgment \\
Release notes & Doc. gen. & Tier 2 & D: Established; V: High; C: Medium \\
Migration script & Code gen. & Tier 1 & C: High (prod data); D: Emerging \\
\bottomrule
\end{tabularx}
\end{table}

API endpoints estimated at 5 story points each (including validation), down from 8 without AI. The ``savings'' of 3 points is real but smaller than the na\"ive assumption that the team used in month one.

\subsection{Execution}

Day~3: A reviewer catches an API endpoint with correct authentication but incorrect business logic. Syntactically valid, tests pass, looks professional---but the business rule contradicts the specification. Tier~2 success: supervision worked.

Day~7: Sampling data from autonomous unit tests (Tier~3) shows two sampled tests correct. But during integration, an un-sampled test has a false-positive assertion. Tier~3 quality signal.

Day~9: A developer admits generating a configuration file and committing it without review---a registry violation. The output is correct, but the process breach is noted.

\subsection{Review and Retrospective}

Findings: API endpoints 67\% first-pass acceptance (2/3); unit test error detected outside sampling; documentation required significant revision; release notes accepted on first review.

Decisions: unit test sampling rate increased from 15\% to 25\% (not a tier demotion---error was caught during integration). Documentation D~rating downgraded to Unproven; next sprint: Tier~1. Process violation discussed without blame; visual indicator added to board. One developer reports: \textit{``I haven't written a unit test from scratch in three months.''} Team schedules human-only testing cycle (Principle~4).

\subsection{What This Illustrates}

The framework operates under realistic conditions: imperfect information, delivery pressure, process violations, quality surprises, early competence erosion. Zero additional roles. Overhead: $\sim$30~minutes classification in planning, existing code review with hybrid checklist, 15~minutes hybrid retrospective.

\section{Confronting the Hard Questions}
\label{sec:hard}

\subsection{What Happens When AI Is Better Than the Reviewer?}

The framework does not require reviewers to be more capable than the AI at \textit{producing} outputs. It requires they can \textit{verify} them---a different, generally easier cognitive task. A developer who cannot write an optimal sorting algorithm can verify one works correctly. When verification exceeds the reviewer's competence, escalation or upskilling is required. If no one can verify, the task should not be delegated at that tier.

\textbf{Honest limitation:} Cases where AI outstrips any available human's verification ability will arise. The response is verification tooling (automated testing, formal verification), not performative human review.

\subsection{Who Validates the Validator?}

Three mechanisms: (1)~periodic human-only execution provides ground-truth competence data; (2)~cross-validation audits at organizational level---external reviewers assess whether validation is effective; (3)~error injection testing for mature implementations---known errors introduced to measure reviewer detection rates.

\textbf{Honest limitation:} These add overhead and may be the first practices sacrificed under pressure---precisely when most needed.

\subsection{Is the Framework Obsolete by Design?}

The decision model does not specify which tasks AI can do; it specifies how to assess and manage delegation based on observed evidence. When capabilities change, evidence changes, and tiers adjust. The principles---accountability, governed delegation, proportional validation, competence maintenance---describe how to manage uncertainty about non-human contributors, independent of AI capability levels.

\textbf{Honest limitation:} Specific thresholds and parameters are informed estimates requiring empirical calibration.

\subsection{Won't Teams Just Ignore This Under Pressure?}

Some will. The framework produces early visible value: better estimates (from the first sprint that budgets validation), reduced post-delivery defects (5--10x correction cost avoided), and clear accountability (eliminating blame-assignment energy). The framework cannot force adoption; it can make the costs of \textit{not} adopting it visible and measurable.

\subsection{Does the Task Model Match How People Actually Use AI?}
\label{subsec:coprod}

HAIF models AI use as discrete delegation: specify, delegate, receive output, validate. This maps well to code generation, report drafting, and data transformation. However, an increasing proportion of AI use is \textit{continuous and conversational}: a professional works alongside an AI iteratively, co-producing output through sustained dialogue. There is no single ``AI output'' to validate.

\textbf{The tension:} When a consultant uses an AI assistant over 40~turns of dialogue to develop a strategy, the result is neither ``human-produced'' nor ``AI-generated'' in any clean sense. Under HAIF, this maps to Tier~1 (Assisted)---technically correct but underspecified regarding what validation means.

\textbf{Preliminary protocol for continuous co-production.} While a full model requires research not yet available, we propose three minimal practices for sustained human--AI co-production sessions:

\begin{enumerate}
\item \textit{Re-grounding checkpoints.} At regular intervals (every 25--30~minutes or upon any significant change of direction), the human pauses and formulates---in their own words, without AI assistance---what they have decided and why. If they cannot articulate the reasoning independently, the AI may be driving the analysis rather than supporting it.

\item \textit{Provenance logging of pivots.} When the AI suggests a direction change that the human adopts, it is noted. At the end of the session, the human reviews these pivot points and consciously evaluates whether each was adopted on its merits or through path-of-least-resistance acceptance.

\item \textit{Adversarial self-check.} Before finalizing the output, the human explicitly identifies the three strongest counterarguments to their conclusions. If they cannot do so without AI assistance, this signals potential over-reliance on the AI's framing.
\end{enumerate}

These practices do not constitute a full tier model for co-production. They are interim measures---a form of cognitive hygiene for a usage pattern that the framework does not yet model comprehensively. Developing a rigorous model for continuous co-production, distinguishing \textit{directive} co-production (human maintains clear control) from \textit{generative} co-production (AI substantially influences direction), is a priority for future work.

\textbf{Honest assessment:} This is a genuine structural limitation. The discrete delegation model covers the most visible cases. It does not yet adequately model continuous collaboration where human and machine contributions are fluid.

\section{Comparison with Existing Frameworks}
\label{sec:comparison}

HAIF is a complementary layer. Table~\ref{tab:comparison} specifies interfaces.

\begin{table}[H]
\centering
\caption{HAIF's relationship with existing frameworks.}
\label{tab:comparison}
\small
\renewcommand{\arraystretch}{1.3}
\begin{tabularx}{\textwidth}{L{1.8cm} L{3cm} L{3.5cm} L{4.2cm}}
\toprule
\textbf{Framework} & \textbf{Relationship} & \textbf{Interface} & \textbf{Complementarity} \\
\midrule
Scrum & Protocol layer & Backlog metadata, DoD, retro & Cadence + coordination; HAIF adds hybrid governance \\
DevOps/SRE & Parallel governance & Error budgets $\to$ tier thresholds & System health; HAIF adds output quality \\
MLOps & Up/downstream & Drift $\to$ tier reassessment & Model lifecycle; HAIF adds output lifecycle \\
AI Gov. & Implementation & Oversight req. $\to$ tier restrictions & Policy constraints; HAIF adds workflows \\
HITL & Extended & Checkpoints $\to$ workflow oversight & Concept; HAIF adds operating system \\
\bottomrule
\end{tabularx}
\end{table}

\section{Implications}
\label{sec:implications}

\subsection{For Project Management Practice}

Planning must account for the full lifecycle cost of AI-assisted work. AI-assisted work is not ``faster human work''---it is a different effort profile. Velocity metrics should disaggregate human, AI, and collaborative contributions. Project managers need new competencies: realistic AI capability assessment, delegation boundary design, validation workload management, and cognitive dependency detection.

\subsection{For Organizational Design}

Organizations need hybrid work governance structures: communities of practice, centers of excellence, or cross-team learning forums. Role descriptions, performance criteria, and career paths must account for hybrid work competencies.

\subsection{For Training and Education}

The distinction between \textit{AI literacy} (using tools) and \textit{AI-critical literacy} (evaluating outputs, detecting failures, maintaining independent judgment) is central. Business schools and certification programs should integrate hybrid team management into PM and organizational behavior curricula.

\subsection{For Research}

HAIF provides defined variables for empirical investigation: tier levels, transition events, error rates, validation effort, competence maintenance cycles, estimation accuracy. Priority directions: longitudinal adoption studies, threshold calibration, competence erosion measurement, controlled estimation accuracy experiments.

\section{Limitations and Future Work}
\label{sec:limitations}

This paper presents HAIF through DSR activities 1--4 and partial analytical evaluation (activity~5). Full empirical evaluation has not been conducted.

\textbf{Scope.} The framework assumes structured work management practices (Agile, Scrum, Kanban). Section~\ref{subsec:beyond} sketches adaptation to Kanban and non-software contexts; these are illustrative, not validated. Informal environments are untested.

\textbf{Parameter calibration.} Threshold recommendations---cycle counts, sampling rates, error bounds---are reasoned estimates requiring empirical calibration. The sampling discussion deliberately avoids unjustified formulas; teams should adopt SQC methods \citep{Montgomery2019} as data accumulates.

\textbf{Discrete delegation model.} As discussed in Section~\ref{subsec:coprod}, the framework models AI use as discrete task delegation. It does not adequately address continuous co-production. Preliminary practices are proposed; a comprehensive model is future work.

\textbf{Multi-agent scenarios.} HAIF deliberately models AI as a single delegation target per task, with a human accountability nexus. It does not address agent-to-agent delegation chains or multi-agent orchestration. This constraint is a design choice, not an oversight. In safety-critical and regulated domains, accountability must resolve to a natural person; diffusing it across an agent chain creates exactly the accountability void that Principle~2 exists to prevent. As agentic architectures mature, extending the delegation model to multi-agent configurations is a priority for future work---but the extension must preserve, not dilute, the human accountability anchor. We expect HAIF's tiered structure to generalize: each agent in a chain would require its own tier classification relative to its human supervisor, creating a hierarchical accountability graph rather than a flat delegation.

\textbf{Organizational and cultural factors.} HAIF presupposes a minimally functional organizational context: teams that conduct regular retrospectives, management that respects quality gates, and an organizational culture where raising concerns about AI output quality is not penalized. Political dynamics, incentive structures, and leadership commitment are critical adoption factors that the framework does not model. In organizations where delivery pressure systematically overrides quality discipline, or where incentive structures reward velocity metrics over correctness, HAIF's protocols will be circumvented regardless of their design quality. This is not unique to HAIF---any process framework assumes a minimum threshold of organizational willingness to follow it---but it is worth stating explicitly: HAIF provides the \textit{what} and \textit{how} of hybrid team governance, not the \textit{why should we bother}, which remains a leadership and change management challenge.

\textbf{Tool specificity.} Tool-agnosticism ensures broad applicability but limits implementation guidance. Companion guides for specific platforms are planned.

\textbf{Cognitive load.} Even minimal protocol overhead adds load for small teams. The adoption friction/rigor trade-off has not been empirically assessed.

\textbf{Future work} should prioritize: (1)~longitudinal case studies of adoption; (2)~quantitative calibration of tier thresholds; (3)~controlled experiments comparing performance with and without HAIF; (4)~extension to multi-agent configurations; (5)~development of a co-production model with cognitive science grounding; and (6)~empirical investigation of the theoretical constructs proposed by \citet{Eccles2025}---particularly cross-species shared mental models and epistemic safety---within HAIF-structured teams.

\section{Conclusion}
\label{sec:conclusion}

Hybrid human--AI teams are an operational reality that existing frameworks do not adequately address. Organizations make delegation, accountability, and quality decisions through improvisation, creating inconsistency, risk, and invisible competence erosion.

HAIF proposes a structured response: four principles, a formal delegation decision model, tiered autonomy with quantifiable transitions, and a protocol layer that integrates into Agile and Kanban workflows without adding roles for small teams. The framework confronts the limits of human oversight, the reality of skill atrophy, the tension between validation discipline and delivery pressure, and its own structural limitations regarding continuous co-production and multi-agent scenarios.

HAIF does not claim to be definitive. It provides a formal operational language for decisions teams already make informally, with mechanisms that make consequences visible and measurable. Its value will be determined empirically---by teams that adopt it, practitioners who refine it, and researchers who test it.

The need is immediate. The response should be equally so.

\newpage
\appendix
\section{Validation Checklists by Task Type}
\label{app:checklists}

These checklists are starting points for Tier~2 (Supervised) validation. Teams should adapt them. The same checklists apply to Tier~3 sampled outputs.

\subsection{Code Generation}

\begin{table}[H]
\small
\renewcommand{\arraystretch}{1.25}
\begin{tabularx}{\textwidth}{L{2.5cm} L{4.5cm} L{5.5cm}}
\toprule
\textbf{Check} & \textbf{What to verify} & \textbf{Common AI failure modes} \\
\midrule
Functional correctness & Does it do what the spec requires? Test including edge cases. & Passes basic tests, fails on boundaries, nulls, concurrency. \\
Business logic & Correct business rules, not just technically valid? & Plausible but incorrect interpretation of requirements. Most dangerous: looks right. \\
Security & Input validation, auth, injection, secrets, dependencies. & Omits hardening, uses deprecated libraries, includes placeholders. \\
Error handling & Failure paths handled? Exceptions caught? Messages informative? & Happy-path implementations. Generic or absent error handling. \\
Performance & Unnecessary loops, redundant queries, N+1 patterns? & Prioritizes readability over performance. Works in dev, fails at scale. \\
Integration & Follows existing patterns, conventions, architecture? & Internally consistent but conflicts with codebase conventions. \\
Hallucinated APIs & All called functions real and available? & Confidently calls non-existent functions or deprecated signatures. \\
\bottomrule
\end{tabularx}
\end{table}

\subsection{Document Generation}

\begin{table}[H]
\small
\renewcommand{\arraystretch}{1.25}
\begin{tabularx}{\textwidth}{L{2.5cm} L{4.5cm} L{5.5cm}}
\toprule
\textbf{Check} & \textbf{What to verify} & \textbf{Common AI failure modes} \\
\midrule
Factual accuracy & Trace every claim to a source. & Fabricated statistics, non-existent studies, misattributions. \\
Source verification & Do cited sources exist and say what's claimed? & Hallucinated citations. Never assume a reference is real. \\
Logical coherence & Does the argument flow? Conclusions supported? Contradictions? & Individually reasonable paragraphs; circular or unsupported overall argument. \\
Context fit & Appropriate for audience, org context, purpose? & Generic content missing client-specific factors. \\
Completeness & Critical topics missing? Scope adequate? & Covers what it can generate fluently; omits what it cannot. \\
Tone/register & Style appropriate and consistent? & Register shifts, generic corporate language, inconsistency. \\
Numerical consistency & Numbers add up? Percentages match absolutes? & Individually plausible numbers that are internally inconsistent. \\
\bottomrule
\end{tabularx}
\end{table}

\subsection{Data Analysis and Transformation}

\begin{table}[H]
\small
\renewcommand{\arraystretch}{1.25}
\begin{tabularx}{\textwidth}{L{2.5cm} L{4.5cm} L{5.5cm}}
\toprule
\textbf{Check} & \textbf{What to verify} & \textbf{Common AI failure modes} \\
\midrule
Input integrity & Correct data processed? All records? Silent losses? & Silently drops non-conforming records, truncates datasets. \\
Transform logic & Logic correct? Test on known pairs. & Off-by-one dates, incorrect joins, wrong aggregation granularity. \\
Statistical validity & Methods appropriate? Assumptions met? & Applies methods without checking assumptions. \\
Edge cases & Missing values, outliers, duplicates handled? & Default handling without flagging decisions. \\
Output format & Matches schema and destination requirements? & Format mismatches causing downstream failures. \\
Reproducibility & Re-runnable with same results? & Non-deterministic ops, undocumented parameters. \\
\bottomrule
\end{tabularx}
\end{table}

\subsection{General Validation Principles}

\textbf{The plausibility trap.} The most dangerous outputs look right. Extra scrutiny for polished, complete-seeming outputs.

\textbf{Verify, don't read.} Reading and finding it ``looks good'' is not validation. Check claims against sources, run code against tests, compare numbers against raw data.

\textbf{Check what's missing.} AI failure often manifests as omission: the risk not mentioned, the edge case not handled, the confound not addressed.

\textbf{Document the review.} Record what was checked, found, and accepted. This feeds feedback loops and enables tier transitions.

\newpage
\section{Quick Start Implementation Guide}
\label{app:quickstart}

This appendix provides a step-by-step protocol for teams adopting HAIF. It assumes a Scrum-based team of 3--7 people already using AI informally. Total additional overhead: approximately 45 minutes per sprint.

\subsection{Before Your First Sprint (One-Time Setup, ~1 Hour)}

\begin{enumerate}[leftmargin=*, itemsep=2pt]
\item \textbf{Assign the Hybrid Work Owner (HWO) role.} This is not a new person. The Scrum Master or Tech Lead absorbs this function. Responsibility: facilitate classification, track tier performance, flag skill erosion.
\item \textbf{Agree on the tier definitions with the team.} Print or display the four tiers (Section~6.4). Everyone must understand what each tier means in terms of who does what and what validation looks like.
\item \textbf{Create a shared tier registry.} A simple table---spreadsheet, Confluence page, or board column---with columns: \textit{Task Type}, \textit{Current Tier}, \textit{Owner}, \textit{Evidence}, \textit{Last Reviewed}. This is the team's institutional memory of what AI can and cannot be trusted with.
\item \textbf{Select validation checklists.} Use Appendix~A as starting points. Adapt to your domain. The checklist doesn't need to be perfect; it needs to exist so that ``review'' means something specific rather than ``read it and see if it looks okay.''
\end{enumerate}

\subsection{Sprint Planning (+30 Minutes)}

For each backlog item:

\begin{enumerate}[leftmargin=*, itemsep=2pt]
\item \textbf{Classify the tier.} Every item gets a classification: Tier~1 (default---human works, AI assists), Tier~2--4 (AI produces output requiring specific validation), or AI-restricted (AI explicitly prohibited for this task). Most items will be Tier~1. The classification makes AI use visible; without it, delegation happens silently and unmanaged. Mark the tier visibly on the board (e.g., a color tag or icon).
\item \textbf{For AI-involved items, assign a tier.} Use the four decision inputs (Section~6.1):
  \begin{itemize}[itemsep=1pt]
    \item[\textbf{S}] \textit{Structuredness:} Are inputs and expected outputs well-defined?
    \item[\textbf{V}] \textit{Verifiability:} Can you objectively check if the output is correct?
    \item[\textbf{C}] \textit{Consequence:} What happens if the output is wrong?
    \item[\textbf{D}] \textit{Demonstrated capability:} How has AI actually performed on this specific task type for your team?
  \end{itemize}
  If the team has no prior evidence for this task type, start at Tier~1 or Tier~2. Never start at Tier~3 or above without documented evidence from lower tiers.
\item \textbf{Assign a named human owner.} This person is accountable for the output regardless of who---or what---produced it. No owner, no delegation.
\item \textbf{Budget validation time explicitly.} Do not estimate only the generation time. Tier~2 items require review time comparable to 30--60\% of original human effort. Add this to the sprint capacity calculation. This is the most predictable integration risk: teams budget for generation but not validation, then either skip reviews or miss the sprint.
\end{enumerate}

\subsection{During the Sprint}

\begin{enumerate}[leftmargin=*, itemsep=2pt]
\item \textbf{Owners validate according to tier.} Tier~1: normal human work, AI assists. Tier~2: full review against checklist before delivery. Tier~3: sample and automated checks. Use the checklists from Appendix~A.
\item \textbf{Record validation outcomes.} Did it pass first review? What errors were found? Where? This data feeds tier transition decisions. Without it, tiers are just guesses.
\item \textbf{Any team member can demote a tier immediately.} If someone spots a pattern of errors, a near-miss, or an AI failure mode not previously seen, the tier drops. No approval process, no waiting for retro. Raise it in standup.
\item \textbf{Flag process violations.} If someone commits AI output without the required review: note it. Don't blame---but do track it. Process violations under pressure are the leading indicator of framework erosion.
\end{enumerate}

\subsection{Sprint Retrospective (+15 Minutes)}

Add these questions to your existing retrospective:

\begin{enumerate}[leftmargin=*, itemsep=2pt]
\item \textbf{Were tiers accurate?} Review the validation records. Did Tier~2 items actually need full review, or were they rubber-stamped? Did any item surprise us with quality problems?
\item \textbf{Was validation effort correctly estimated?} Compare budgeted vs.\ actual validation time. Adjust estimation baseline for next sprint.
\item \textbf{Any tier promotions or demotions?} Promotion requires: minimum 3 consecutive sprint cycles at current tier with documented quality data. Present the evidence; team decides.
\item \textbf{Skill erosion check.} Ask explicitly: is anyone losing the ability to do work they're now delegating? If a developer hasn't written a certain type of code in months, schedule a human-only cycle for that task type next sprint (Principle~4).
\item \textbf{Update the tier registry.} Record decisions and rationale.
\end{enumerate}

\subsection{Common Mistakes in the First Three Sprints}

\begin{itemize}[leftmargin=*, itemsep=2pt]
\item \textbf{Starting too many items at Tier~3.} Teams overestimate AI reliability based on impressive demos. Start conservative (Tier~1 or~2) and promote with evidence.
\item \textbf{Not budgeting validation time.} The sprint fills up with generation tasks, review becomes a bottleneck, shortcuts follow. Budget validation as a first-class activity.
\item \textbf{Assigning ownership pro forma.} If the ``owner'' never actually reviews the output, ownership is performative. The owner must have time, skill, and authority to reject.
\item \textbf{Treating the classification as bureaucracy.} If the team sees this as overhead rather than risk management, it will be abandoned by sprint~3. The HWO's job is to make classification feel like a natural part of planning, not a compliance exercise.
\item \textbf{Ignoring the skill erosion signal.} It's invisible until it's critical. If nobody on the team can write a unit test without AI, you cannot validate AI-generated tests. Schedule human-only work before this becomes an emergency.
\end{itemize}

\subsection{Minimum Viable Adoption}

For teams that want to start with the absolute minimum:

\begin{enumerate}[leftmargin=*, itemsep=2pt]
\item Tag every AI-produced deliverable with a human owner's name.
\item Before the owner approves, they check it against the relevant checklist (Appendix~A).
\item In the retrospective, ask: ``Did AI help or create extra work this sprint?''
\end{enumerate}

That is three practices. Everything else in HAIF is refinement. But these three---ownership, structured validation, and honest retrospection---address the core risks. Start here, add the tier model when the team is ready.

\newpage
\bibliographystyle{apalike}

\end{document}